# Is the ultra-fast transformation of bismuth non-thermal?

E. G. Gamaly, A. V. Rode

*Laser Physics Centre, Research School of Physics and Engineering*
*Australian National University, Canberra, ACT0200, Australia*

Abstract

Transient state of femtosecond laser excited bismuth has been studied by various groups with time-resolved optical, x-ray, and electron probes at the deposited energy density from below through up to several times the equilibrium enthalpy of melting. However, the interpretations of the experimental results are controversial: the optical probes reveal the absence of transition to the melting phase while the authors of x-ray and electron diffraction experiments claim the observation of ultrafast non-thermal melting. The presented analysis, based on temperature dependence of bismuth optical properties, unequivocally shows a purely thermal nature of all the observed fs-laser induced transformations in bismuth.

## I. INTRODUCTION

The advent of ultra-fast optical, x-ray, and electronic diagnostic techniques have provided an insight into such most common phase transformation as crystal melting at atomic space and time scales. The transient state of the femtosecond laser excited Bi crystal has been studied with time-resolved fast optical [1-10], x-ray [11-13] and electron [14,15] probes at the deposited energy density below and up to eight times above the equilibrium enthalpy of melting. However, there is no consensus in interpretation of the experiments with the same crystal in nearly identical excitation conditions. The fast fall-off of x-ray probe intensity reflected from bismuth crystal [12] as well as the intensity decrease for the diffracted electron beam [14] were interpreted as the evidence of non-thermal lattice melting induced solely by the electronic excitation, while the dual optical probe of bismuth reflectivity in the same conditions clearly demonstrate the absence of transition to melting state and presents an evidence of purely thermal character of the ultra-fast material transformation [7,8].

A recent publication of Sciaini *et al.* [14] is another striking justification of the existing controversy. The authors have examined the transient states of the 30-nm *freestanding* bismuth films irradiated by 200-fs laser pulses through femtosecond electron diffraction. It is remarkable that the sharp decrease in the diffracted intensity was observed even at the absorbed energy density half of the equilibrium melting enthalpy when the disordering due to melting is energetically impossible. Obviously, the reason for the observed intensity decrease calls for a different explanation.

The measured temperature dependence of optical properties of bismuth in equilibrium conditions in the temperature range from room temperature up to 773 K (~215K over the melting point) [7,10,17-19] allows one obtaining the high temperature properties of Bi directly from the experiments. The material properties recovered from the optical data, such as the electron-phonon momentum and the energy transfer rates, the heat diffusion coefficient, the number of electrons in the conduction band and others, form a strong basis for the quantitative interpretation of the experimental results with the ultra-fast excitation of bismuth. The time for setting up the statistical distributions and the electron-lattice thermalisation time also directly follow from the optical data. The diffusion coefficient recovered from the optical data is in close proximity to that obtained from x-ray reflectivity in fs-excited bismuth experiments by Johnson *et. al.* [13].

The structure of the paper is as the following: (i) – we retrieve the temperature dependencies of the electron-phonon momentum transfer rate and the plasma frequency (the number of electrons in the conduction band) obtained from the dielectric function dependency on temperature in the equilibrium conditions using the Drude-like form for the dielectric function of bismuth; (ii) – we derive theoretically the temperature-dependent electron-phonon momentum exchange rate and its link to the energy exchange rate. We fit the momentum rate to the experimental data and then establish the temperature-dependent energy exchange rate corresponding to the experiments; (iii) – we estimate the absorbed laser energy and the lattice temperature in Bismuth layer excited by laser with fluences (0.5 – 23) mJ/cm$^2$; (iv) – we compare the electron-phonon energy transfer time to the characteristic time of decrease in the electron and x-ray diffracted intensity to the quasi-stationary level as the function of the absorbed laser fluence; (v) – we identify the processes, which may induce non-homogeneous heating of the skin-layer, which, in turn, are the reasons for the breaking of the Bragg conditions and therefore for the observed diffracted intensity decrease. Finally, we draw the conclusion on the nature of transformation in bismuth excited by the fs-laser pulse. The analysis presented in this paper unequivocally shows a purely thermal nature of all observed transformations in bismuth excited by ultra-fast laser pulses.

## II. TEMPERATURE-DEPENDENT BISMUTH PROPERTIES FROM OPTICAL EXPERIMENTS IN EQUILIBRIUM

### A. Electron-phonon momentum exchange rate

The measurements of the reflectivity and the dielectric function by the ellipsometry technique, which gives simultaneously the real and imaginary parts of dielectric function allow direct recovering the electron density and the electron-phonon momentum exchange rate. Indeed, the reflectivity, *R,* is directly related to the real $\varepsilon_r$ and imaginary $\varepsilon_i$ parts of the dielectric function, $\varepsilon = \varepsilon_r + i\varepsilon_i$ through the Fresnel formula:

$$R = \left|\frac{\sqrt{\varepsilon}-1}{\sqrt{\varepsilon}+1}\right|^2 = \frac{|\varepsilon| + 1 - \sqrt{2(|\varepsilon| + \varepsilon_r)}}{|\varepsilon| + 1 + \sqrt{2(|\varepsilon| + \varepsilon_r)}}; \qquad (1)$$

where $|\varepsilon| = \sqrt{\varepsilon_r^2 + \varepsilon_i^2}$. It was found long ago [17-20] that the dielectric function for solid Bi at room temperature and that for the liquid Bi obeys to the Drude-like form. The dielectric function in the Drude form is directly linked to the plasma frequency $\omega_p$ and the electron-phonon momentum exchange rate, $\nu_{e\text{-}ph}$:

$$\varepsilon_r = 1 - \frac{\omega_p^2}{\omega^2 + \nu_{e-ph}^2}; \quad \varepsilon_i = \frac{\omega_p^2}{\omega^2 + \nu_{e-ph}^2}\frac{\nu_{e-ph}}{\omega}; \qquad (2)$$

Here $\omega$ is the laser field frequency, $\omega = 2.356\times10^{15}$ s$^{-1}$ for 800 nm, and $\omega_p^2 = 4\pi e^2 \frac{n_e}{m_e^*}$, $e$ is the electron charge and $m_e^*$ is the effective electron mass. Thus the electron-phonon momentum exchange rate and plasma frequency, measured in units of laser frequency, are directly connected to the real and imaginary parts of the dielectric function:

$$\frac{\nu_{e-ph}}{\omega} = \frac{\varepsilon_i}{1-\varepsilon_r}; \quad \frac{\omega_p^2}{\omega^2} = (1-\varepsilon_r)\left(1 + \frac{\nu_{e-ph}^2}{\omega^2}\right). \qquad (3)$$

If the electron mass is known than the number density of the conductivity electrons is directly retrieved from the plasma frequency. Following this procedure Garl [7] recovered the following Bi properties from the ellipsometry measurements at the room temperature $T_{RT}$ = 294K for laser light of 800 nm: $\frac{\omega_p^2}{\omega^2}$ = 31; $R$ = 0.74; $\nu_{e\text{-}ph} = 2.1\times10^{15}$ s$^{-1}$; $n_e = 5.34\times10^{22}$ cm$^{-3}$; Fermi energy $\varepsilon_F$ = 5.17 eV ($v_F = 1.35\times10^8$ cm/s) under assumption that electron has a free electron mass. Comins [17] found that Bi at 773 K has the following parameters: The reflectivity is $R$ = 0.67; $(\omega_{pe}/\omega)^2$ = 81.58 that corresponds to the number density of free carriers $n_e = 1.42\times10^{23}$ cm$^{-3}$. Therefore all 5 valence electrons are transferred into conduction band. Comins suggested that electron has a free mass value [17], then the Fermi energy $\varepsilon_F$ = 9.92 eV, $v_F = 1.87\times10^8$ cm/s in molten Bi. The electron-phonon momentum exchange rate reads $\nu_{e\text{-}ph} = 5.67\times10^{15}$ s$^{-1}$. Therefore, the available data from 293 K to 773 K changing seemingly smoothly. However, there is a phase transition point from solid to liquid state, the melting point, at 544.7 K. It is known that Bi, along with ice, Sb, Ga, KI, and InSb, belongs to so-called open crystalline structure [21] so its molten phase density is higher than the solid state. Little is known about the behaviour of other physical parameters of Bi near the phase transition point from the literature. Moreover, there is no consensus, to the best of our knowledge, if the melting of Bi is the first order or the second order phase transition. As it follows from the text below, physical parameters recovered from the optical experiments such as number density of electrons in the conduction band and the Fermi energy value, also apparently smoothly changing with the temperature. We have no data near and above the melting point, so we decided to interpolate the available data by smooth continuous dependence on the temperature as a first approximation until some new data become available. Some support to this action comes also from the theory presented further down in the paper, where the numerical coefficients in the linear temperature dependence for the electron-phonon momentum exchange rate coincide with the experimental values and are almost identical for both solid and liquid states. Therefore, linear interpolation of the data extracted from all available experiments [7,17,18], and keeping in mind the theory (see in Section III.A below) shows that $\nu_{e\text{-}ph}$ grows up in direct proportion to temperature (Fig.1):

$$\nu_{e-ph} = 2.1\times10^{15}\frac{T}{T_{RT}}\left[s^{-1}\right]. \tag{4}$$

The linear dependence of the momentum exchange rate on temperature fits the optical measurements for bismuth in equilibrium with sufficient accuracy; the proportionality to the temperature holds for Bi well before and long after the melting point [7,17-19].

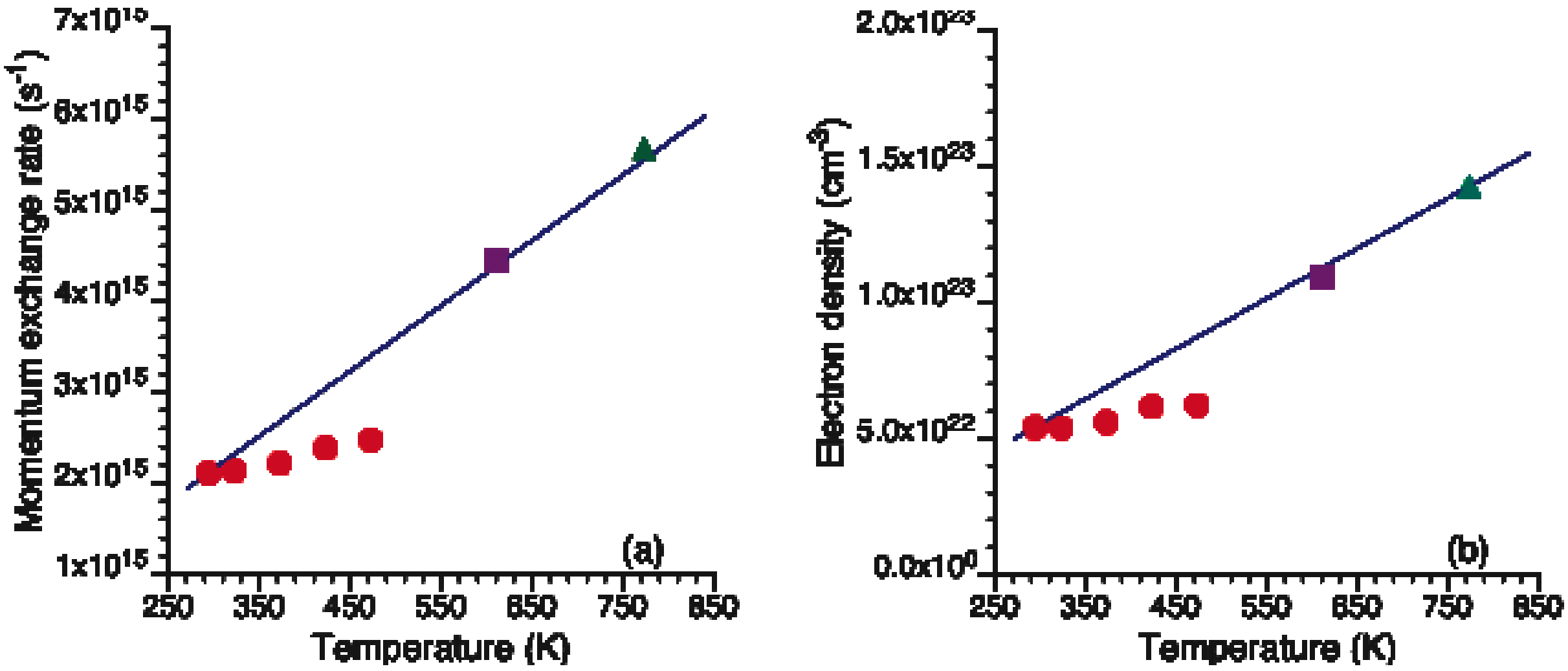

Fig.1. Temperature dependence of the momentum exchange frequency (a), and of the electron density (b), both in equilibrium conditions; solid line corresponds to linear dependence Eq.(3,4). The circles are the results from ellipsometry measurements Ref.[7], triangle – Ref.[17], square – Ref.[18].

## B. Number of electrons in the conduction band

The electrons number density in the conduction band is retrieved from the plasma frequency under assumption that the electron mass is equal to a free electron mass: $m^* = m_e$. The result at room temperature gives $n_e = 5.34\times10^{22}$ cm$^{-3}$ (1.89 out of total 5 valence electrons are in the conduction band) and $\varepsilon_F = 5.17$ eV (Fig.1b). At 773 K in liquid bismuth all five electrons are in the conduction band: $n_e = 1.42\times10^{23}$ cm$^{-3}$, $\varepsilon_F = 9.92$ eV, and the dielectric function obeys the Drude-like form [10,17,18]. Thus the experiments had shown that from 40 to 100% of the valence electrons are transferred to the conduction band at the temperature increase from the room temperature to the melting point. These data are in a sharp contrast to the statements in [11,12,14] that only from 5 to 15% of valence electrons are in the conduction band under the strong laser excitation to a much higher temperature.

## C. Electronic heat conduction and characteristic cooling time

Thermal diffusivity (coefficient of thermal diffusion) relates to both the Fermi energy and the electron-phonon momentum exchange rate: $D = v_F^2 / 3\nu_{e-ph}$. The diffusivity values recovered from the temperature dependencies in equilibrium [10,17-19], are $D$ = (2.0 – 2.89) cm$^2$/s for the temperature range 293 K – 793 K. These results are in good agreement with the recent non-equilibrium measurements from the x-ray reflectivity data of fs-laser excited bismuth giving $D$ = 2.3 cm$^2$/s [13]. Note the drastic difference of this value from 0.067 cm$^2$/s given for equilibrium conditions in the reference book [22]. Accordingly, the electron mean free path, $l_{mfp} = \frac{v_F}{\nu_{e-ph}}$ = 0.67 nm << $l_s$, is much less than the skin layer depth and the film thickness, assuring the legitimacy of the diffusion approximation for the electron heat transfer.

The time for the temperature smoothing across a 30 nm thick film [14] taken in accord to the diffusivity value is, respectively, $t_{smooth} = l_s^2 / D$ = 3.9 ps, which is a reason for preserving a high thermal gradient across the film thickness in the first ps after the laser excitation.

## III. LINK BETWEEN TEMPERATURE-DEPENDENT ELECTRON-PHONON MOMENTUM AND ENERGY EXCHANGE RATES: THEORY

### A. Electron-phonon momentum exchange rate

The electrons are travelling in a periodic self-consistent potential $W(r)$ of an ideal crystalline lattice without losses. The electron-phonon scattering is the main source of changes in the electron momentum and energy. The potential slightly perturbed ($W_0 >> \delta W(r)$) by the electron-phonon scattering, this could be presented as an expansion of $W(r)$ into series of small perturbations [23]:

$$W(r) = W_0 + \delta W(r) + \delta W''(r) + ... \equiv H_0 + H'_{e-ph}(r) + H''_{e-ph}(r) + ... \qquad (5)$$

The first correction to the potential corresponds to the emission or absorption of a phonon with quasi-momentum to be conserved. The perturbation of the potential expresses through the amplitude of atomic vibrations (phonons amplitude) Δ, as the following:

$$\delta W(r) = \sum_j \left\lfloor \frac{\partial W(r)}{\partial r} \right\rfloor_{0j} \Delta_j \qquad (6)$$

The summation over the all phonon states is assumed. With the help of quantum-mechanical perturbation theory one can find the probability of electron transition from a state marked by indices, ($p$,$l$), to a state ($p'$,$l'$), simultaneously absorbing or emitting a phonon $\hbar\omega_{kj}$ in the $k$,$j$ mode [23]:

$$w_{p'l';pl} = \frac{2\pi}{\hbar}\left|H'_{p'l';pl}\right|^2 \delta\left(\varepsilon_{pl} - \varepsilon_{p'l'} - \hbar\omega_{kj}\right); \tag{7}$$

Here the delta-function represents the energy conservation. On the classic (as oppose to quantum) language the above probability is the electron-phonon momentum exchange rate.

Let us estimate the perturbation Hamiltonian as $H' \cong \frac{\partial W}{\partial r}\Delta \cong \frac{J_0\Delta}{d}$; the value of the potential is of the order of the first ionisation potential $J_0$, $W \cong J_0$ ($J_0$ = 7.3 eV for bismuth), while the space scale of the potential gradient is of the order of magnitude of the inter-atomic distance $d$. The electron energy is significantly larger than the phonon energy $\hbar\omega_{ph}$ and it is practically unchanged in a collision, $\varepsilon_F >> \hbar\omega_{ph}$. Then, Eq.(7) converts into the general expression for the electron-phonon momentum exchange rate $\nu_{e-ph}^{mom}$:

$$\nu_{e-ph}^{mom} \approx w_{p'l';pl} = 2\pi \frac{J_0^2}{\hbar\varepsilon_F}\frac{\Delta^2}{d^2}. \tag{8}$$

At the temperature below the Debye temperature, $T_D$, the electron-phonon momentum exchange rate (and the phonon's amplitude) does not depend on the lattice temperature, ($T_D$ = 119 K for bismuth) $k_B T \le \hbar c_s k \approx \hbar\omega_D$, here $c_s$ is the speed of sound, $k$ is the wave vector, and $\omega_D$ is the Debye frequency [23-25]. The phonon amplitude is $\Delta^2 \approx 2\hbar / M\omega_D$. Then, taking into account that $M\omega_D^2 d^2 \approx \varepsilon_b$, one obtains the collision frequency to be proportional to the Debye frequency, $\nu_{e-ph}^{mom} \cong 4\pi \frac{J_0^2}{\varepsilon_F\varepsilon_b}\omega_D$, at low lattice temperature.

At the higher temperature $T_L > T_D$ one should average the squared phonon amplitude (energy) in Eq.(8) over the Maxwell-Boltzmann distribution. For 1D oscillator this average expresses as the following [26,27]:

$$\left\langle \Delta^2 \right\rangle = \frac{\int_{-\infty}^{\infty} \Delta^2 \exp\left(-M\omega^2\Delta^2 / 2k_B T_L\right) d\Delta}{\int_{-\infty}^{\infty} \exp\left(-M\omega^2\Delta^2 / 2k_B T_L\right) d\Delta} = \frac{k_B T_L}{M\omega^2}.$$

For harmonic oscillator in three dimensions the average squared phonon's amplitude reads:

$$\left\langle \Delta^2 \right\rangle \approx \frac{3k_B T_L}{M\omega_{ph}^2}. \tag{9}$$

The phonon frequency conventionally expressed through the second derivative of the inter-atomic potential in equilibrium:

$$M\omega_{ph}^2 \approx \left(\frac{\partial^2 U}{\partial r^2}\right)_0 \approx \frac{\varepsilon_b}{d^2}; \tag{10}$$

$\varepsilon_b$ = 2.16 eV for bismuth. Now Eq.(8) with the help of Eqs.(9,10) reads:

$$\nu_{e-ph}^{mom} \approx 2\pi a \frac{J_0^2}{\varepsilon_F \varepsilon_b} \frac{k_B T_L}{\hbar} . \tag{11}$$

For Bi: $J_0$ = 7.3 eV; $\varepsilon_F$ = 5.17 eV, therefore the material-dependent coefficient in Eq.(11) $2\pi \frac{J_0^2}{\varepsilon_F \varepsilon_b}$ = 29.98, while the numerical coefficient $a$ = (1-3) can be used for fitting to the experimental data. For the room temperature 294 K one gets for solid Bi from Eq.(11) $\nu_{e-ph}^{mom}(T_{RT})$ = (1.15-3.46)×10$^{15}$ s$^{-1}$. Thus, theory gives this rate proportional to the temperature in agreement with the optical ellipsometry measurements [7,17-20]. It is interesting that the experimental value of $\nu_{e-ph}$ = 2.1×10$^{15}$ s$^{-1}$ ($\nu_{e-ph}/\omega$ = 0.89) at room temperature could be attributed to the electron interaction with a two-dimensional harmonic oscillator at $a \cong 2$.

## B. Electron-phonon energy exchange

The electron-phonon energy exchange includes the processes of emitting and absorbing phonons. Therefore the second term in the expansion of the interaction Hamiltonian should be taken into account:

$$H''_{e-ph} = \delta W''(r) \approx \frac{1}{2}\left(\frac{\partial^2 W}{\partial r^2}\right)_0 \Delta^2 \approx \frac{1}{2}\frac{J_0}{d^2}\Delta^2 \tag{12}$$

The electron-phonon energy exchange rate now expresses as the following:

$$\nu_{e-ph}^{en} \approx w''_{p'l';pl} = \frac{2\pi}{\hbar}\left|H''_{p'l';pl}\right|^2 \delta\left(\varepsilon_{pl} - \varepsilon_{p'l'} - \hbar\omega_{kj}\right) \approx \frac{\pi J_0^2}{2\hbar \varepsilon_F}\frac{\Delta^4}{d^4} \tag{13}$$

The electron-phonon energy exchange rate (the inverse of the electron – lattice temperature equilibration time) is linked to the electron-phonon momentum exchange rate Eq.(8) as the following:

$$\nu_{e-ph}^{en} \cong \nu_{e-ph}^{mom} \frac{\Delta^2}{4d^2}. \tag{14}$$

At the temperature below $T_D$ it reads:

$$\nu_{e-ph}^{en} \approx \frac{2\pi J_0^2}{\hbar\varepsilon_F\varepsilon_b}\frac{\hbar\omega_D}{\varepsilon_b}\omega_D; \qquad (15)$$

That is in a qualitative agreement to [28]. The Eq.(14) provides a basis for obtaining the energy exchange rate from the electron-phonon momentum exchange rate directly extracted from the optical experiments. One shall take the ratio of the squared phonon's amplitude to the inter-atomic distance in Eq.(14) corresponding to the two-dimensional oscillator as suggested by the experimental data $\langle\Delta^2\rangle/d^2 = 2k_BT_L/\varepsilon_b$. Then Eq.(14) takes the form:

$$\nu_{e-ph}^{en} \cong \nu_{e-ph}^{mom}\frac{k_BT_L}{2\varepsilon_b}. \qquad (16)$$

Now, taking the link between the temperature dependent energy exchange and the collision frequency and using the experimental temperature dependence Eq.(4) one obtains the electron-phonon energy exchange rate dependence on temperature as the following:

$$\nu_{e-ph}^{en} \cong 10^{15}\frac{k_BT_{room}}{\varepsilon_b}\left(\frac{k_BT}{k_BT_{room}}\right)^2 = 1.2\times10^{13}\left(\frac{k_BT}{k_BT_{room}}\right)^2\left[s^{-1}\right]. \qquad (17)$$

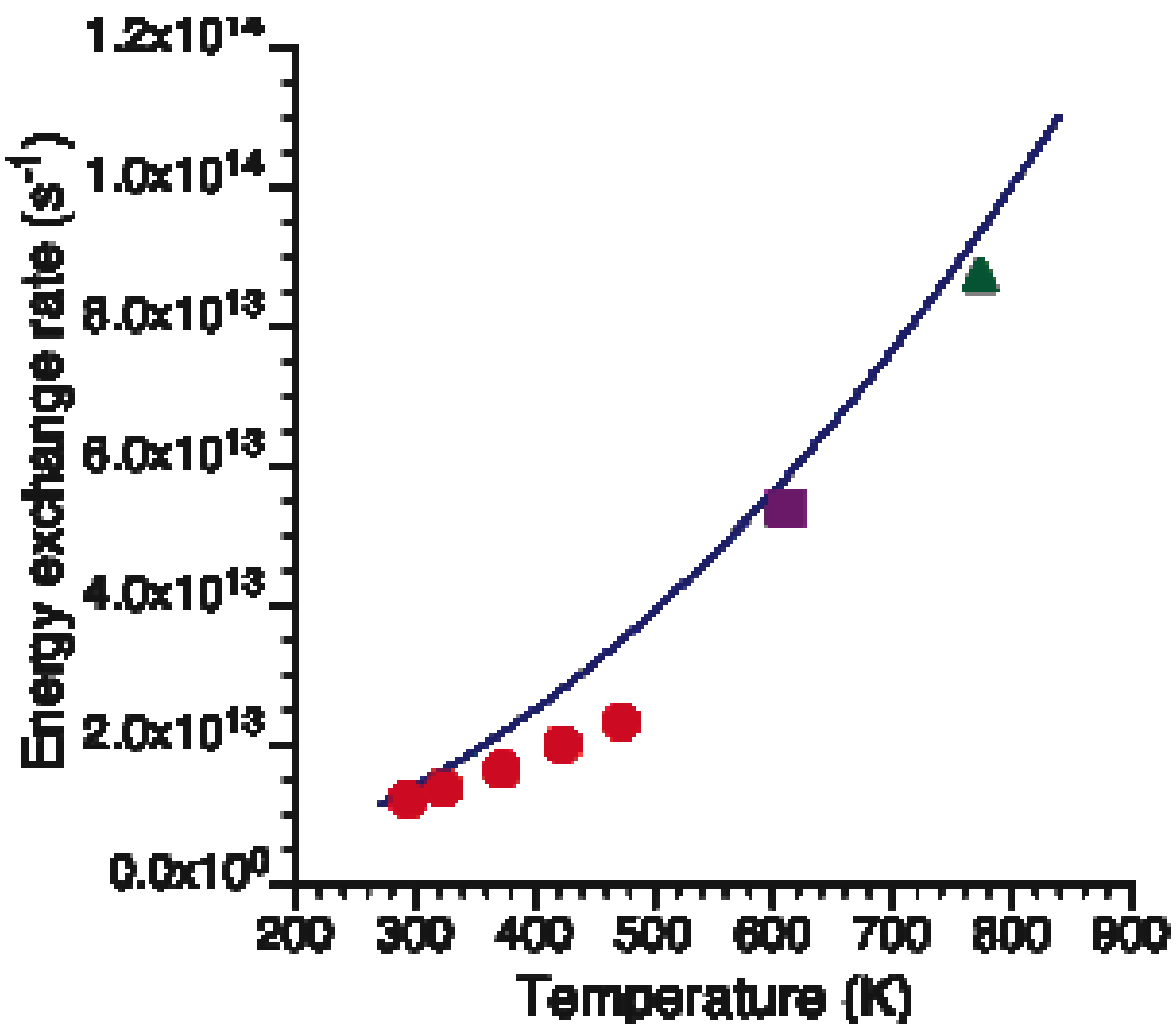

Fig.2. Energy exchange rate: solid line corresponds to Eq.(17). Circles are the results from ellipsometry measurements Ref.[7], triangle – Ref.[17], square – Ref.[18]

The results of calculations of the energy exchange rate (with correction taken from experiments) are compared in Fig.2 to the available experimental data. Now it is also possible calculating the dielectric function and reflectivity changes with temperature, which are presented in Fig.3. The dielectric function at room temperature was measured to be $\varepsilon_r = -16.25$, $\varepsilon_i = 15.4$ [7,10] at 800 nm; the values are very close to the literature data [20,22]. The electron-phonon collision rate and plasma frequency at room temperature were found with the help of Eqs.(2,3) as $\nu_{e-ph}^{mom} = 0.893\omega$ and $\omega_{pl}^2 = 31.0\omega^2$. The temperature dependencies were calculated with the help of Eqs.(11,17). The theory predicts well the experimental results of dielectric function measurements for liquid bismuth at 610 K [18] and 773 K [17].

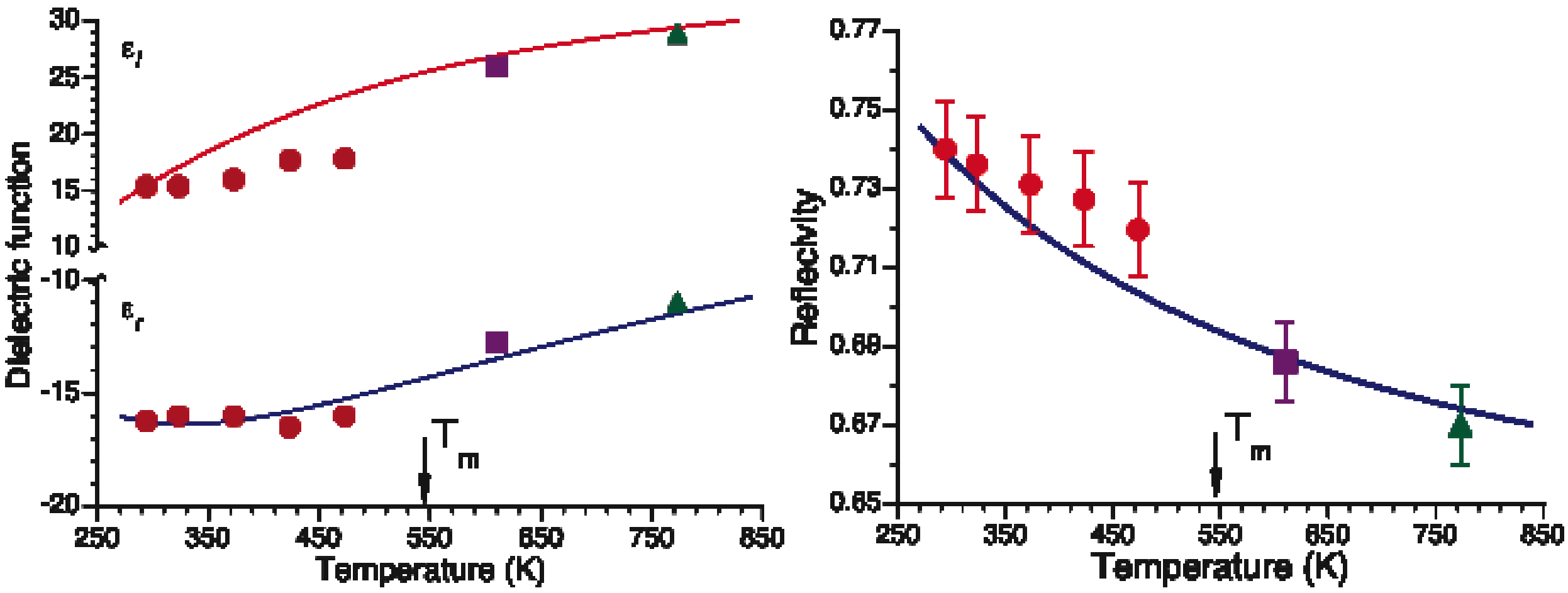

Fig.3. Temperature dependencies of real $\varepsilon_r$ and imaginary $\varepsilon_i$ parts of the dielectric function (left) and reflectivity (right) of bismuth at 775 nm in equilibrium conditions. The solid lines were calculated using Eqs.(1,2,17); the circles are results of ellipsometry measurements from Ref.[7], triangles – Ref.[17], squares – Ref.[18], arrows indicate the melting point $T_m = 544.7$K.

## IV. ULTRAFAST LASER INTERACTION WITH BISMUTH

### A. Absorbed laser energy, temperature and heat conduction

The absorbed laser energy density $E_{abs}$ in a skin layer excited by a fs-laser pulse is defined by the following relation [29]:

$$E_{abs} \cong \frac{2AF(t_p)}{l_s}; \quad F(t_p) = \int_0^{t_p} I(t)dt, \tag{18}$$

where $I(t)$ is laser intensity during the pulse of duration $t_p$. A laser beam with the wavelength 775 nm excites Bismuth in the skin layer $l_s$ = 28 nm with the absorption coefficient $A$ = 0.26 [9,10]. The laser-excited Bismuth was studied in the fluence range, $F(t_p)$= (0.5 – 23) mJ/cm$^2$ [9-12] that corresponds to the absorbed energy density range from 0.13 kJ/cm$^3$ to 4.3 kJ/cm$^3$, i.e. from three times below through to eight times higher the equilibrium enthalpy of melting of 0.53 kJ/cm$^3$ (that corresponds to the incident fluence of 2.85 mJ/cm$^2$). The statistical distribution in a sub-system of identical particles is much faster than the laser pulse, providing a basis for the two-temperature description of the laser-excited matter [30,31,28]. The electron-electron collisions are responsible for establishing distributions in electron sub-system (electron temperature) [26], while the phonon-phonon collisions lead to the establishing the lattice temperature [23]. The absorbed energy density (intensity) and therefore the temperature exponentially decay inside the skin layer as the follows:

$$T(x,t) = T_{\max}(t)\exp\{-2x/l_s\} \tag{19}$$

The temperature has a maximum at the vacuum-sample interface. The electron temperature decreases when electrons transfer their energy to the lattice via electron-phonon collisions and due to heat conduction later in time. As it follows from Fig.2 the energy exchange time for the studied range of laser fluences is shorter than the pulse duration. Therefore the electron-lattice equilibration occurs early in the pulse time and temperature ($T_e$=$T_L$) reaches maximum at the end of the pulse. The maximum lattice temperature after the electron-lattice energy equilibration explicitly links to the absorbed laser energy due to the energy conservation:

$$T_L \cong \frac{E_{abs}}{(C_L n_a + C_e n_e)}, \tag{20}$$

where $C_L$, $C_e$ are the lattice and electron heat capacity correspondingly, and $n_a$, $n_e$ are the atomic and electron density in solid Bi [9,10]. We assume here that $C_L = 3k_B$ and $C_e = \frac{\pi^2 k_B T_e}{2\varepsilon_F}$, $k_B$ = 1.38×10$^{-23}$ J/K is the Boltzmann constant. The maximum temperature dependencies on the laser fluence are presented in Fig.4.

The electron temperature at the vacuum-sample interface of the film is much lower than the Fermi energy $\varepsilon_F$ = 5.17 eV even for the maximum fluence of 23 mJ/cm$^2$. Therefore the Fermi electrons with velocity $v_F$ = 1.35×10$^8$ cm/s are responsible for the heat transport across the film.

Thus, the temperature at the outer film boundary decreases due to the linear electronic heat conduction in accord to:

$$T_{\max}(x=0;t) \approx T_{\max}(x=0;t_{cool})\left(\frac{t_{cool}}{t_{cool}+t}\right)^{1/2} \qquad (21)$$

The cooling time for skin layer with the thickness $l_s$ is introduced above in the conventional form:

$$t_{cool} = l_s^2 / D; \qquad (22)$$

here $D$ is the coefficient of thermal diffusion, which is linked directly to the electron velocity, $v_F$ (the Fermi velocity), and electron-phonon momentum exchange rate, $\nu_{e\text{-}ph}$, by the relation:

$$D = v_F^2 / 3\nu_{e-ph}. \qquad (23)$$

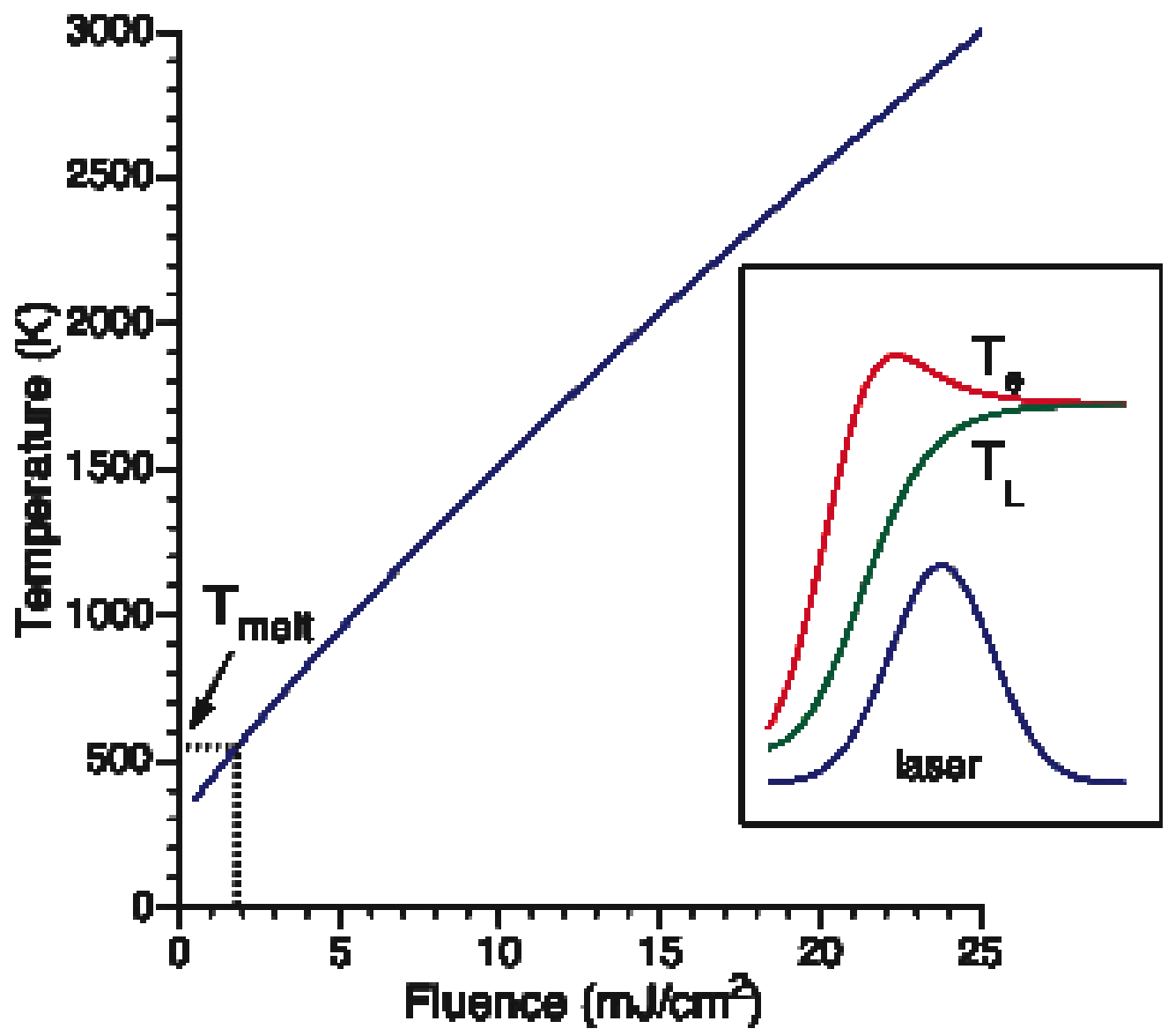

Fig. 4. Maximum lattice temperature after the electron-lattice energy equilibration *vs* incident laser fluence. The inset shows typical electron and lattice temperature behaviour during the laser pulse in the case when the energy equilibration time is shorter than the fs-laser pulse.

B. Energy exchange time vs diffraction intensity fall-off in fs-laser excited bismuth

The dependence of the energy exchange time $t_{e-ph}^{en} = \left(\nu_{e-ph}^{en}\right)^{-1}$ for the broad range of fluences in the laser-excited Bi is presented in Fig.5 and compared to the measured characteristic time for the diffraction intensity decrease in the experiments [11,12,14]. Analysis of the data from Fig.5 allows drawing at least two conclusions. First, it is remarkable that the intensity of the diffracted beam decreases significantly even at the fluences when the absorbed energy density is lower than the equilibrium enthalpy of melting. Therefore the melting is not the reason for the violation of the Bragg conditions and the intensity decrease. The laser-beam induced non-homogeneities in the excited sample might be responsible for the observed decrease. We discuss those later in the paper. The second conclusion from the analysis of Fig. 5 is that the times for the diffracted intensity decrease are always much longer than the electron-phonon energy exchange time, which represents the lattice heating time. Hence it follows, that the observed intensity decay is thermal in nature.

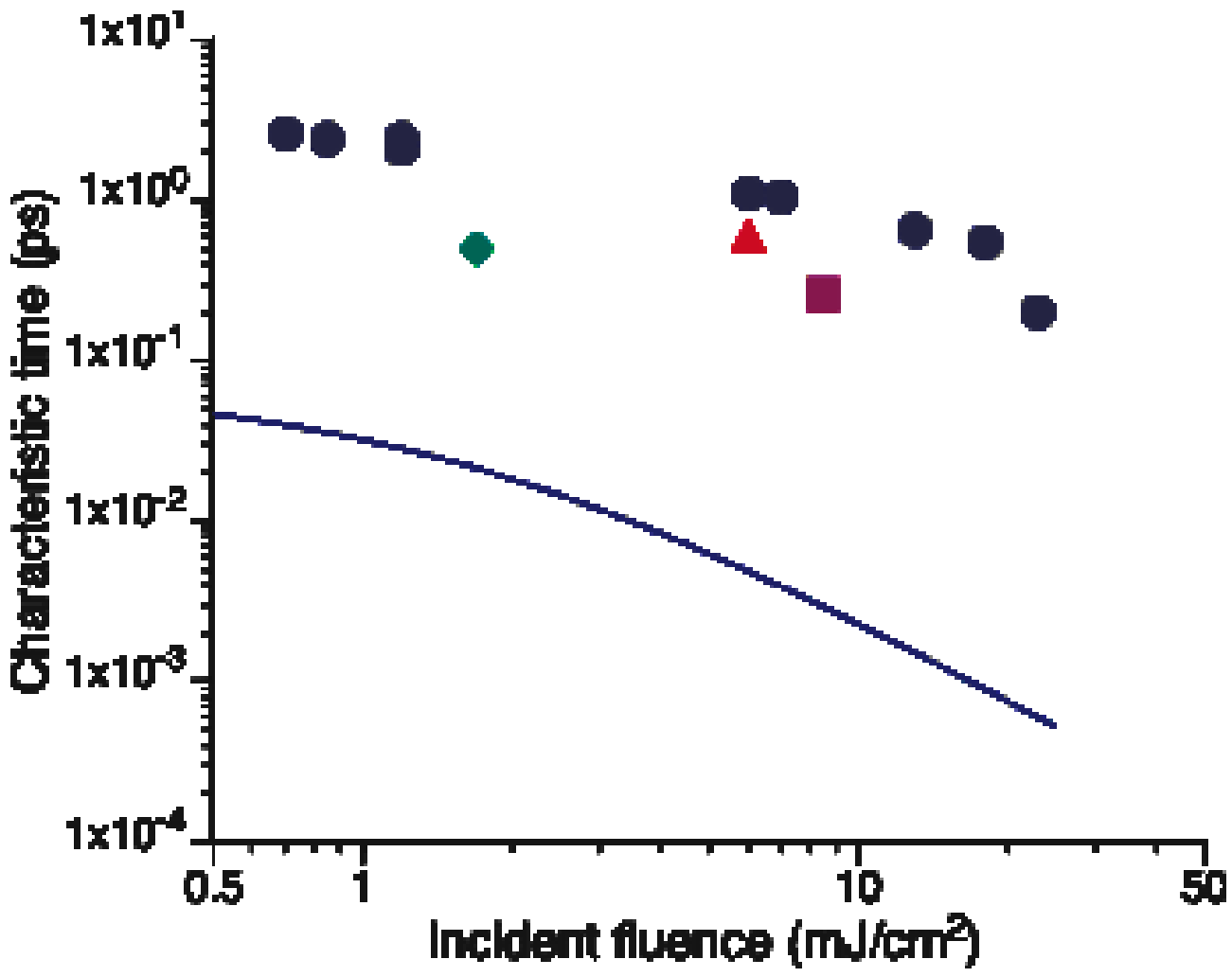

Fig.5. Electron-lattice energy exchange time $t_{e-ph}^{en} = \left(\nu_{e-ph}^{en}\right)^{-1}$ according to Eq.(17) (solid line) and characteristic decrease time of the diffraction intensities: circles are from the electron diffraction experiments Ref.[14]; diamond (Ref.[11]), triangle (Ref.[12]), and square (Ref.[13]) are from x-ray diffraction experiments.

At the low fluence range the absorbed laser energy is up to five times lower than the enthalpy of melting. Therefore melting could not be the reason for the fall off of the diffraction intensity. The energy transfer from electrons to the lattice and the diffraction intensity fall-off occurs simultaneously at low intensity, while at superheating above the enthalpy of melting the thermalization process is much faster. Thus all processes at the absorbed energy density at and above the enthalpy of melting occur in conditions when electron and lattice temperature are equilibrated and the main part of the Maxwell-Boltzmann distribution (temperature) is already established. Therefore all the transformations occurred are of the thermal nature. We discuss below the spatial anisotropy induced by the laser beam in the skin layer that might be a reason for the violation of the Bragg's conditions.

C. Laser-induced spatial inhomogeneity across the laser-excited film

Let us discuss the possible sources for the laser-induced spatial inhomogeneities in the laser-excited sample ignored in the previous analyses. The spatial distribution of the absorbed laser energy across the skin layer might be the primary source of the observed fast drop of the diffracted beam intensity. Indeed, the absorbed laser energy decreases exponentially in the $l_s$ = 28 nm bismuth skin layer: $E_{abs}(x) = E_{abs,\max}\exp(-2x/l_s)$ – see Fig.6. Therefore, both the absorbed energy density and temperature have maximum at the vacuum-sample interface irradiated by the laser pulse. The surface atoms start to expand after the energy transfer from the electrons with the speed of sound in Bi of $1.79\times10^5$ cm/s. Thus in a picosecond time the outer interface moves on 17.9 Angstroms, more than 4 inter-atomic distances. Note that the gradient force $F = -\nabla(T_e + T_L) = (T_e + T_L)/l_{skin}$ acts on atoms inside the skin layer slightly moving them in direction of the laser beam (into the sample) – Fig.6. Definitely a coherent displacement affects the probe beam diffraction. There is no obvious way of distinguishing the decrease in the diffraction beam intensity caused by the inhomogeneity from that in the thermally disordered media.

The laser interaction with a freestanding layer in [14] imposes further implications for interpretation of the electron diffraction pattern. The pressure at the front surface of the film, which is equal to the absorbed energy density, is ~35 kBar at 23 mJ/cm$^2$, while at the rear surface it is ~4 kBar. The sound velocity in solid Bi is $1.79\times10^5$ cm/s; thus the front surface of the free standing film shifts towards the laser beam during the 200 fs pulse by 3.6 Å while the rear surface expands in the opposite direction. This creates additional source for the intensity decrease in the diffracted electron beam irrelevant to melting. Note that the temperature smoothing time due to heat conduction across the skin layer (see the end of Section II.C) is 3.9 ps, which is much longer than the time for the diffracted intensity fall-off in Fig. 5. Possible inhomogeneous laser intensity distribution of both the pump and the probe beams over the surface induce further complication into the interpretation of the diffracted beam intensity behaviour.

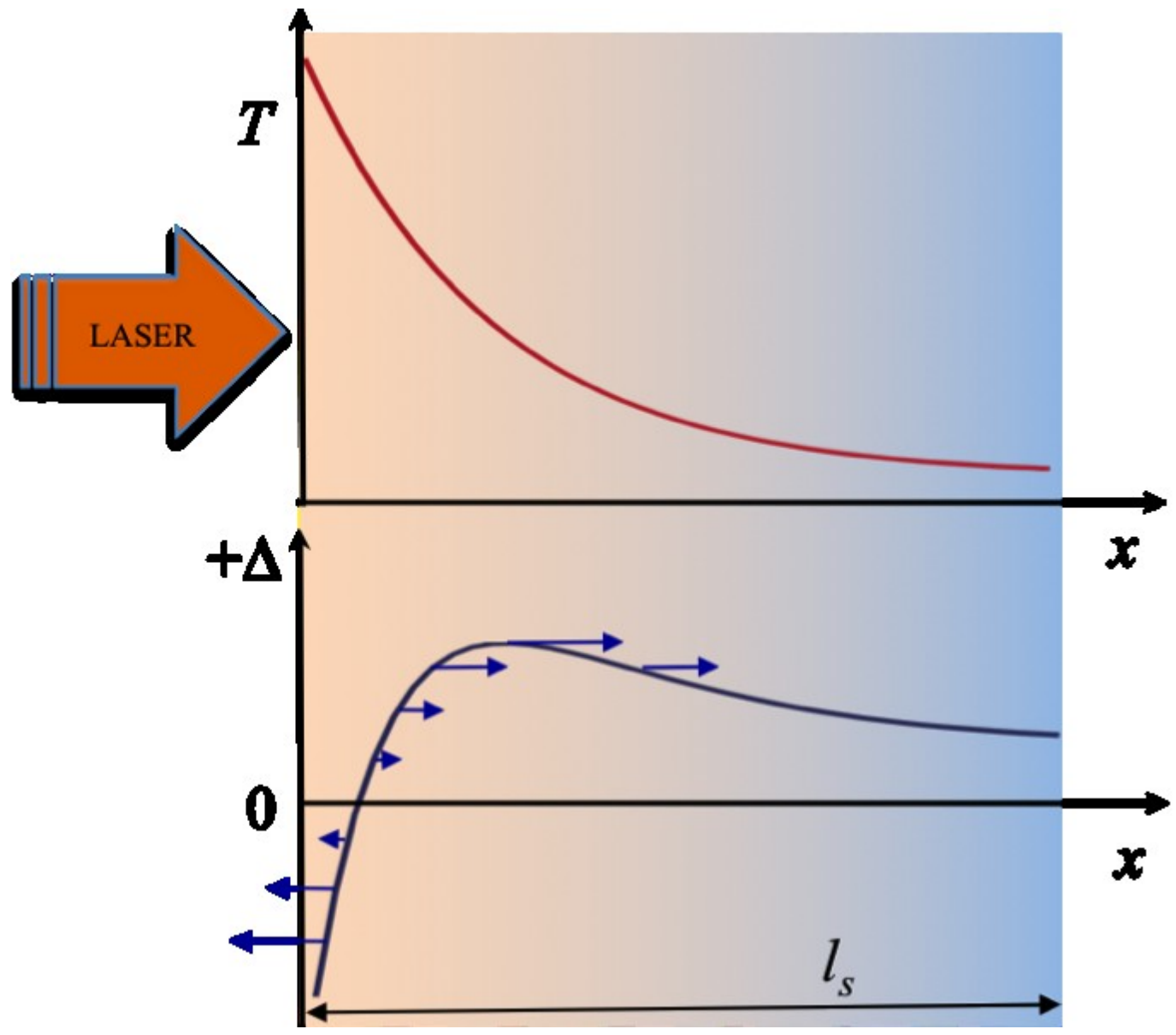

Fig.6. Schematic representation of exponential temperature decrease (top graph) with depth in the skin layer, and the atomic displacement Δ along the *x*-axis (lower graph) due to thermal pressure (negative shift toward the laser beam) and the gradient pressure (positive along the laser beam); $l_s$ = 28 nm for 800 nm wavelength in bismuth.

D. The diffracted intensity decrease in the ultra-fast experiments and its relation to the classic Debye-Waller factor

Many authors [12,14,15] use the classic Debye-Waller factor for determining the lattice temperature in a solid swiftly excited by the ultra-fast laser. Let us discuss the validity of classical (equilibrium) approach at the temperature approaching the melting point or in the case when atomic displacements are of *"well-directed motion"*. The decrease in the beam intensity diffracted from a medium where the atoms are displaced at a small ($u \ll G^{-1}$) distances $u$ expresses through atomic displacement and the reciprocal lattice vector $G$ [24]:

$$\frac{I}{I_0} = \left\langle \exp\left(-iG \cdot u\right) \right\rangle \cong 1 - i\left\langle G \cdot u \right\rangle - \frac{1}{2}\left\langle \left(G \cdot u\right)^2 \right\rangle - \ldots \qquad (25)$$

The angular brackets means averaging over the distribution function. The condition $\left\langle G \cdot u \right\rangle$ = 0 holds if $u$ is a random displacement uncorrelated with $G$. That applies to the thermal harmonic vibrations, $\langle u \rangle \equiv \int u f_B du$, distributed in accord with the Boltzmann function $f_B \propto \exp\left\{-M\omega_{ph}^2 u^2 / k_B T\right\}$, here $\omega_{ph}^2$ is the constant phonon frequency. At the temperature approaching the melting point the atomic vibrations lose their harmonic character. Then, the third order correction to the interaction potential becomes significant, $\Delta U \approx M\omega_{ph}^2 u^2 - g u^3$. Now the average displacement from the equilibrium position (the first term in RHS of Eq.(25)) becomes non-zero, $\langle u \rangle \approx \frac{k_B T}{\varepsilon_b} d$ [10,28=24] while the dependence of the average phonon energy on the temperature still holds the same form as for the harmonic vibrations, $\frac{1}{2} M\omega_{ph}^2 \langle u^2 \rangle \approx \frac{3}{2} k_B T$ (here the harmonic oscillations are in three dimensions [24]). Therefore at the temperature approaching the melting point the classic Debye-Waller expression, ignoring the first term in Eq.(25), becomes invalid. The first and the second order terms of the diffracted intensity decay are both non-zero in respect to the averaged displacement and are proportional to the lattice temperature. The first order term is also non-zero in the case of coherent atomic displacement induced by the electronic pressure gradient. For these reasons the classic Debye-Waller expression cannot be used to determine the lattice temperature at the elevated temperature close to the melting point.

## V. DISCUSSION AND CONCLUSIONS

We would like emphasising the several important issues missed in the interpretation of the experiments with Bismuth swiftly excited by ultrafast laser pulses.

First, the recovery of the temperature dependence for the electron-phonon momentum and energy exchange rates, for the number density of the conductivity electrons, and thermal diffusion at the temperatures up to 200K above the melting point allows performing the quantitative interpretation of the experiments with the ultra-fast excitation of Bi.

The number of electrons in the conduction zone from the published data [7,18-20] at the temperature in excess the room temperature ranges from 40% to 100% of the valence electrons. Modification of the potential surface by the electron excitation occurs in Bi in a way similar to that in other metals such as aluminium [16] or gallium [32], and can be related to the increase in the inter-atomic spacing due to directed, coherent displacement [9-11]. It is worth noting that coherent displacement is rather similar to thermal expansion.

Secondly, the temperature dependence of the electron-phonon energy exchange time allows unequivocally establish purely thermal nature of the ultra-fast laser induced transformations in Bismuth: the electron-phonon temperature setting up and energy equilibration occurs faster than the transformation takes place, that is in accord with the observed transformations in aluminium [16] and gallium [32].

Thirdly, the laser-induced inhomogeneity in the absorbed energy density and temperature results in the significant atomic displacements, comparable to the inter-atomic distance, might be a major source for the violation of the Braggs condition and the diffraction intensity decrease when the electron and x-ray probes were used.

The additional concern in interpretation of laser-induced melting in Bi relates to the fact that liquid bismuth in equilibrium conditions is denser than solid. In the typical experimental set up for the ultra-fast excitation the temperature is a maximum at free vacuum-sample boundary, therefore the importance of expansion is obvious. A special diagnostics is needed to distinguish the effect of expansion on the diffracted probe from the phase transition changes in a sample.

We also demonstrate that the use the classical Debye-Waller factor for the interpretation of the experiments in non-equilibrium conditions should be revised: the first order term in the atomic displacement at a temperature close to the melting point (that is zero for harmonic vibrations) is of the same order of magnitude as the second-order term. The use of the classical factor for the defining the temperature is questionable.

One should note that the decrease in the phonon frequency when the temperature approaches the melting point signifies the onset of the vibration instability [25,26,33,34]. This is the first in a succession of instabilities preceding melting [34]. The frequency decrease relates to the lattice reconstruction and it does not indicate the termination of atomic vibrations.

To conclude we have shown that the fast fs-laser induced excitation of Bi in all experiments published so far indicate the entirely thermal nature of the observed transformations.